\documentclass[12pt,a4paper]{article}
\usepackage{amsmath,amssymb}
\usepackage{hyperref}
\usepackage[numbers,compress]{natbib}

\makeatletter
\font\manfnt=manfnt
\def\Watchout{\@ifnextchar [{\W@tchout}{\W@tchout[1]}}
\def\W@tchout[#1]{{\manfnt\@tempcnta#1\relax%
  \@whilenum\@tempcnta>\z@\do{%
    \char"7F\hskip 0.3em\advance\@tempcnta\m@ne}}}
\def\@W@tchout#1{\W@tchout[#1]}
\def\dubious{\@ifnextchar[{\@dubious}{\@dubious[1]}}

\def\@dubious[#1]{%
  \setbox\@tempboxa\hbox{\@W@tchout#1}
  \@tempdima\wd\@tempboxa
  \list{}{\leftmargin\@tempdima}\item[\hbox to 0pt{\hss\@W@tchout#1}]}
\newif\if@preliminary
\@preliminaryfalse
\def\preliminary{\@preliminarytrue}
\def\preprintno#1{\def\@preprintno{#1}}
\def\address#1{\def\@address{#1}}
\def\email#1#2{\thanks{\tt #1@{}#2}}
\def\abstract#1{\def\@abstract{#1}}
\renewcommand\abstractname{ABSTRACT}
\newlength\preprintnoskip
\newlength\abstractwidth
\@titlepagetrue
\renewcommand\maketitle{\begin{titlepage}%
  \let\footnotesize\small
  \hfill\parbox{\preprintnoskip}{%
  \begin{flushright}\@preprintno\end{flushright}}\hspace*{1cm}
  \vskip 60\p@
  \begin{center}%
    {\Large\bf\boldmath \@title \par}\vskip 1cm%
    {\sc\@author \par}\vskip 3mm%
    {\@address \par}%
    \if@preliminary
      \vskip 1cm {\large\sf PRELIMINARY DRAFT}%
    \fi
  \end{center}\par
  \@thanks
  \vfill
  \begin{center}%
    \parbox{\abstractwidth}{\centerline{\abstractname}%
    \vskip 3mm%
    \@abstract}
  \end{center}
  \end{titlepage}%
  \setcounter{footnote}{0}%
  \let\thanks\relax\let\maketitle\relax
  \gdef\@thanks{}\gdef\@author{}\gdef\@address{}%
  \gdef\@title{}\gdef\@abstract{}\gdef\@preprintno{}
}%
\def\@citex[#1]#2{\if@filesw\immediate\write\@auxout{\string\citation{#2}}\fi
  \def\@citea{}\@cite{\@for\@citeb:=#2\do
    {\@citea\def\@citea{,\penalty\@m}\@ifundefined
       {b@\@citeb}{{\bf ?}\@warning
       {Citation `\@citeb' on page \thepage \space undefined}}%
\hbox{\csname b@\@citeb\endcsname}}}{#1}}
\def\citerange{\@ifnextchar [{\@tempswatrue\@citexr}{\@tempswafalse\@citexr[]}}
\def\@citexr[#1]#2{\if@filesw\immediate\write\@auxout{\string\citation{#2}}\fi
  \def\@citea{}\@cite{\@for\@citeb:=#2\do
    {\@citea\def\@citea{--\penalty\@m}\@ifundefined
       {b@\@citeb}{{\bf ?}\@warning
       {Citation `\@citeb' on page \thepage \space undefined}}%
\hbox{\csname b@\@citeb\endcsname}}}{#1}}
\long\def\@makecaption#1#2{%
  \vskip\abovecaptionskip
  \sbox\@tempboxa{#1: \emph{#2}}%
  \ifdim \wd\@tempboxa >\hsize
    #1: \emph{#2}\par
  \else
    \hbox to\hsize{\hfil\box\@tempboxa\hfil}%
  \fi
  \vskip\belowcaptionskip}
\def\fmslash{\@ifnextchar[{\fmsl@sh}{\fmsl@sh[0mu]}}
\def\fmsl@sh[#1]#2{%
  \mathchoice
    {\@fmsl@sh\displaystyle{#1}{#2}}%
    {\@fmsl@sh\textstyle{#1}{#2}}%
    {\@fmsl@sh\scriptstyle{#1}{#2}}%
    {\@fmsl@sh\scriptscriptstyle{#1}{#2}}}
\def\@fmsl@sh#1#2#3{\m@th\ooalign{$\hfil#1\mkern#2/\hfil$\crcr$#1#3$}}
\makeatother

\newcommand{\re}[1]{\mbox{\boldmath$#1$}}
\newcommand{\rb}[1]{\mbox{\boldmath$\overline{#1}$}}
\newcommand{\vev}[1]{\mbox{$\langle#1\rangle$}}
\usepackage{graphicx}
\usepackage{amsmath}
\allowdisplaybreaks
\begin{document}


\baselineskip20pt   
\preprintno{SI-HEP-2014-09\\[0.5\baselineskip] May 2014}
\title{%
 Flavour Models with Three Higgs Generations
}
\author{%
 F.~Hartmann\email{hartmann}{physik.uni-siegen.de},
 W.~Kilian\email{kilian}{physik.uni-siegen.de},
}
\address{%
Universit\"at Siegen, Department Physik,
      57068~Siegen, Germany\\
}
\abstract{%
We construct models with a spontaneously broken $SU(3)_F$ flavour
symmetry where three generations of Higgs multiplets transform in a
flavour-triplet representation.  The models are embedded in a
supersymmetric Pati-Salam GUT framework, which includes
left-right symmetry.  We study the possible flavon representations and
show that a model with flavons in the decuplet representation is able
to reproduce the hierarchy structure of the quark and lepton mass and
mixing matrices.  This result requires no additional structure or
fine-tunings except for an extra $Z_4$ discrete symmetry in the
superpotential.
}
\maketitle

\section{Introduction}
\label{sec:intro}

The Standard Model (SM) contains a single complex doublet of
scalar fields with one physical Higgs particle, which stands out as an
extra field, unrelated to the fermionic matter or gauge fields.
The recent discovery at the CERN Large Hadron Collider (LHC) is a
clear indication that Higgs bosons exist \cite{Aad:2012tfa,Chatrchyan:2012ufa}.
However, with the current data sample, it is not possible to distinguish the
minimal Higgs representation from a larger one, in particular from the MSSM
Higgs sector or from a model that contains more than one Higgs (bi-)doublet.

In this work, we study the particular case that there are three Higgs
generations\cite{Lee:1973iz,Weinberg:1976hu,Grossman:1994jb} which can be
associated with a flavour 
symmetry~\cite{Keus:2013hya,Ivanov:2012fp,Aranda:2012bv}, thus
relating the Higgs and matter sectors.  Such
a structure is present, for instance, in supersymmetric grand-unification models
where Higgs and matter fields combine in a common gauge-group
representation~\cite{Hewett:1988xc,King:2005my}.

Extra Higgs generations contribute to the running of the electroweak gauge
couplings and therefore adjust the conventional picture of gauge
unification.  In Ref.~\cite{Hartmann:2014fya}, we have investigated Pati-Salam
models \cite{Pati:1973uk,Pati:1974yy} and found a considerable number of
models with interesting patterns of gauge unification that contain three Higgs
generations.  Similar results can be found in
refs.~\cite{Arbelaez:2013hr,Braam:2009fi,Kilian:2006hh,Howl:2007hq}.

A flavour symmetry may be either discrete or continuous (for reviews see
\cite{Altarelli:2010gt,King:2013eh,King:2014nza,King:2003jb}). In this paper,
we focus on a continuous $SU(3)_F$ flavour symmetry. There is 
a large number
of models that realize this
idea~\cite{King:2003jb,King:2001uz,King:2003rf,deMedeirosVarzielas:2005ax,
King:2006me,Antusch:2004xd,King:2005bj},
where the $SU(3)_F$ transformations that apply to left- and right-handed, up-
and down-type quarks and leptons respectively, may be either connected or
independent. In a three-generation Higgs model, the Higgs doublets (bi-doublets
in the supersymmetric case) also provide a fundamental $SU(3)_F$ triplet.  We
consider the simplest possibility that a single $SU(3)_F$ symmetry universally
applies to all Higgs and matter multiplets, and assume that it is
spontaneously broken by vevs of additional scalar fields which we collectively
denote as flavons. In order to avoid Goldstone bosons, we assume the 
flavour $SU(3)$ to be gauged. Unlike the Higgs scalars, we do not 
imply that flavons appear at collider energies.\footnote{%
  We are not concerned with the dynamics of flavon fields, but merely treat
  them as spurions that parametrize the possible patterns of flavour-symmetry
  breaking.}

Without further assumptions, one should consider flavons in any
representation of $SU(3)_F$, not just the fundamental triplet representation.
As we will show, models with only triplet flavons are feasible, but a model
with a dominant decuplet contribution appears more natural and is able to
qualitatively accommodate the observed mass- and mixing-hierarchy patterns in
both the quark and lepton sectors.

\section{Model Framework}
\label{sec:framework}

We study extensions of the minimal supersymmetric Standard Model (MSSM) with a 
gauged $SU(3)_F$ flavour symmetry.  The flavour symmetry is 
spontaneously broken, at
high energy, by the vevs of flavon fields~$\phi$.  Both the three generations
of matter and the three generations of Higgs bi-doublets transform in the
fundamental triplet representation of $SU(3)_F$.  At high energies, we 
impose  Pati-Salam symmetry (PS) which contains a discrete left-right
symmetry.  This setup allows for complete unification of the SM gauge couplings 
at or near the Planck scale with a hierarchy of intermediate mass
scales~\cite{Hartmann:2014fya}.

The complete gauge symmetry we start with is thus,
\begin{equation}
 SU(4)_C \times SU(2)_L \times SU(2)_R \times Z_{LR} \times SU(3)_F.
\label{eq:gaugedsymmetry}
\end{equation}
We have to treat this as an effective theory expanded in a large (but
unspecified) mass scale $M$.  In our framework, the flavons are
singlets under the PS gauge group and thus do not contribute to the
corresponding gauge-coupling scale dependence.
As the theory is constructed at a high scale, we also have to take into 
account all non-renormalizable terms which are not forbidden by a remaining 
symmetry or by explicit constraints on the low-energy effective theory after 
symmetry breaking. This allows, in general, mass terms for those components of 
the flavon superfields that are not absorbed as Goldstone bosons into massive 
flavour-gauge bosons.

Generically, we denote the quotient of a flavon vev $\vev{\phi}$ and the
expansion scale $M$ by $\epsilon$,
\begin{equation}
 \epsilon = \frac{\vev{\phi}}{M}\,.
\end{equation}
We assume that $\epsilon$ distinguishes the up- from the down sector,
$\epsilon_d \neq \epsilon_u$, and also allow for a third distinct
expansion parameter $\epsilon_\nu$ in the neutrino sector.  Such
differences are expected, in the low-energy effective superpotential, as a
consequence of the breaking of the Pati-Salam gauge symmetry.

An effective Yukawa-coupling matrix with the entry pattern
\begin{align}
Y_{u/d} \approx  \left( \begin{matrix} 0 				&
O\left(\epsilon^3\right) & O\left(\epsilon^3\right)	\\
					    O\left(\epsilon^3\right)	&
O\left(\epsilon^2\right) & O\left(\epsilon^2\right)	\\
					    O\left(\epsilon^3\right)	&
O\left(\epsilon^2\right) & O\left(1\right)		 \end{matrix}\right)
y_{t/b} \,,
\label{eq:wantedYukmatrix}
\end{align}
with $\epsilon_u \approx 0.05$ and $\epsilon_d \approx 0.15$ can accommodate the
observed CKM matrix and the quark-mass hierarchies~\cite{Roberts:2001zy}.

In a PS model, the matter spectrum necessary contains right-handed neutrinos.
A Majorana mass term for those induces the type-I see-saw
mechanism~\cite{Minkowski:1977sc,Yanagida:1979as,GellMann:1980vs}.  After
integrating out the heavy right-handed neutrinos, the effective mass matrix
for the SM neutrinos takes the form
\begin{equation}
 m_\text{eff} = m_{LR} M^{-1}_{RR} m_{LR}^T\,,
\label{seesawformula}
\end{equation}
where $m_{LR} = Y_\nu v_{SM}$ is the left-right mixing mass matrix from the
lepton-Yukawa terms and $M_{RR}$ is the Majorana mass matrix of the
right-handed neutrinos.  For the Yukawa structure given above,
sequential right-handed neutrino dominance (SRHND)~\cite{King:1999cm}
implements a phenomenologically viable PMNS mixing matrix and mass
hierarchy in the neutrino sector~\cite{King:2003jb}. Here, a single right
handed neutrino gives the main contribution to the 23 block. A second right
handed neutrino is required to give subdominant contributions to generate the
full effective neutrino mass matrix.

\section{A Triplet-Flavon Model}
\label{sec:Tripletmodel}


We first consider the case that all flavons $\phi$ are in the (anti-)
fundamental triplet representation of the flavour symmetry, analogous to
matter multiplets
$\Psi$~\cite{King:2001uz,King:2003rf,deMedeirosVarzielas:2005ax}. The
three generations of Higgs bidoublets $h$ also transform as a flavour
triplet. In order to generate Majorana mass terms for the right-handed
neutrinos, we augment this spectrum by a right-handed PS-breaking multiplet
$\Phi_{R}$, as introduced in~\cite{Hartmann:2014fya}. 
We list the fields and quantum numbers that are relevant for the present 
discussion in Tab.~\ref{tab:fieldcontenttriplets}.

\begin{table}
\centering
\begin{tabular}{cccccc}
\hline\noalign{\smallskip}
Field	& $SU(3)_F$	& PS	& $U(1)$ & $Z_3$		& $Z^R_2$	
\\
\noalign{\smallskip}\hline\noalign{\smallskip}
$\Psi_L$	& $\re{3}$	& $(4,2,1)$	& $1$	& $0$	& $-$ \\
$\Psi_R$	& $\re{3}$	& $(\overline{4},1,2)$ & $1$ & $0$	& $-$ \\
$h$		& $\re{3}$	& $(1,2,2)$	& $1$   & $0$	& $+$ \\[1ex]
$\phi_3$	& $\re{3}$	& $(1,1,1)$	& $-3$	& $-1$	& $+$ \\
$\overline{\phi_3}$	& $\rb{3}$	& $(1,1,1)$	& $-1$	& $0$	& $+$ \\
$\phi_{23}$	& $\re{3}$	& $(1,1,1)$	& $0$	& $1$	& $+$ \\
$\overline{\phi_{23}}$ & $\rb{3}$	& $(1,1,1)$	& $-1$	& $1$	& $+$
\\[1ex]
$\Phi_R$	& $\re{1}$	& $(4,1,2)$ 	& $6$ 	& $0$	& $+$ \\
\noalign{\smallskip}\hline
\end{tabular}
\caption{Transformation of the Higgs and matter and flavon superfields 
responsible for the flavour structure under the gauge and flavour symmetries in 
the flavon triplet model.}
\label{tab:fieldcontenttriplets}
\end{table}

The spectrum is completed by further superfields, restoring left-right 
symmetry as well as canceling the $SU(4)_C$ gauge anomaly, and implement 
symmetry breaking down to the MSSM~\cite{Hartmann:2014fya}. Since those 
superfields do not affect the flavour structure, we do not list them 
explicitly.

In this model, the Pati-Salam and flavour symmetries allow for the SM
gauge couplings and Yukawa terms (see below), but also for extra
flavour-dependent interactions which are inconsistent with the
observed low-energy flavour phenomenology.  We supplement the gauge
structure by a $U(1)\times Z_3$ symmetry and assign specific charges
to the fields, such that unwanted terms are forbidden.  (To
  avoid an extra Goldstone boson, the $U(1)$ may also be considered to be
  gauged, or we may reduce it to a discrete $Z_N$ sub-symmetry.)
These quantum numbers are also given in Tab.~\ref{tab:fieldcontenttriplets}.  
Furthermore, we impose a $Z^R_2$ symmetry similar to matter parity.

The particle content of Tab.~\ref{tab:fieldcontenttriplets}, including 
all PS breaking fields, exhibits a $SU(3)_F$ gauge anomaly. This anomaly can be 
canceled by additional fields that do not contribute to the matter flavour 
structure. For instance, these can be the fields $\overline{\phi}_{10}$ and 
$\phi_6$, which transform as $\rb{10}$ and $\re{6}$, respectively, and are 
PS-singlets. We expect such or similar fields to be introduced in the context 
of vacuum alignment of the flavon vevs \cite{Altarelli:2005yx}.

We label the flavons $\phi_i$ by the directions in flavour space in which
their vevs are aligned.  The flavour symmetry is broken in two steps. First we
break $SU(3)_F$ down to $SU(2)_F$ by the vev of the flavon field $\phi_3$. The
field $\phi_{23}$ then further breaks the remaining flavour symmetry. For the
sake of simplicity we align the ``MSSM-Higgs'' vev along the third component in
flavour space. Collecting all
flavour-breaking Higgs fields, we propose the vacuum structure
\begin{subequations}
\begin{align}
  \vev{\phi_3} \;\sim\; \vev{\overline{\phi}_3} &\sim M\,\left(\;0\;,\;0\;,\;\,
     \sqrt[3]{\epsilon}\;\right) \,, \\
  \vev{\phi_{23}} \sim \vev{\overline{\phi}_{23}} &\sim M
  \left(\;0\;,\;\epsilon\;,\;\,\epsilon\;\right)\,, \\
  \vev{\Phi_R} &\sim \; M_\nu \,,\\
  \vev{h}&=\left(\;0\;,\;0\;,\;1\right)\,v_\text{MSSM} \,.
\end{align}
\label{eq:vevstructure}
\end{subequations}

\subsection{Quark Mixing: Operator Analysis}

We now construct the possible interactions of matter with Higgs and flavon
fields that yield contributions to low-energy flavour physics.  The effective
operators are invariant under Pati-Salam, $SU(3)_F$, and the extra $U(1)$ and
$Z_3$ symmetries, but not limited by the condition of renormalizability.

Starting with only the gauge and flavour symmetry
of~(\ref{eq:gaugedsymmetry}), we identify a single dimension-three
(renormalizable) term in the superpotential~\footnote{We count the dimension
  in terms of the superpotential, where a dimension-three term is
  renormalizable, and terms with mass dimension of four and higher are
  non-renormalizable.},
\begin{equation}
  Y_0 = \varepsilon^{ijk}\; \Psi_{L,i}\,\Psi_{R,j}\,h_k\,,
  \label{eq:trivinvariantstructure}
\end{equation}
where $\varepsilon_{ijk}$ is the three dimensional Levi-Civita-Tensor.
This term generates a coupling that is non-diagonal among the lepton and
quark generations.  Taking a look at (\ref{eq:wantedYukmatrix}), such a term
can not be the dominant contribution to flavour interactions.  It has to be
suppressed at least by a factor $\epsilon^3$ with respect to the top Yukawa
coupling.  We choose to eliminate this term by the
extra $U(1)$ symmetry mentioned above.

We now consider further operators of increasing dimension, one by one.
At dimension four, the algebra of $SU(3)_F$ does not provide an
invariant polynomial that can generate a Yukawa operator.

At dimension five, we again have to forbid all interactions, using the extra
symmetries. These operators generate terms of the form
\begin{align}
 \frac{\overline{\phi}\,\phi}{M^2}\;Y_0,
\end{align}
which, after symmetry breaking, would re-introduce the universal Yukawa term
mentioned above. Its parametric suppression is only of order $\epsilon^2$,
which is not sufficient.

At dimension six, we finally identify an operator that should contribute to
the Yukawa matrices\footnote{The notation is thus, that a pair of fields
  should be contracted
  with a $\delta_i^j$ and likewise a triple with $\varepsilon_{ijk}$},
\begin{equation}
W_\text{lead} =\frac{1}{M^3} (h\; \overline{\phi}_{3})(\Psi_L\, \overline{\phi
}_{3})(\Psi_R\,\overline{\phi }_{3})\,.
\end{equation}
This term generates Yukawa couplings for the third generation.  At the same
order, the model is able to seed the 2-3-block of the Yukawa matrix.  The
corresponding invariant operator is
\begin{equation}
W_{(2,3)}= \frac{1}{M^3} (h\, \overline{\phi
}_{23})(\Psi_L\,\overline{\phi}_{23})(\Psi_R\,\overline{\phi}_{23})\,.
\end{equation}
It is important that, at this order, the universal
term~(\ref{eq:trivinvariantstructure}) is not generated.  As long as there are
only two distinct flavon fields involved, this term vanishes by antisymmetry.

In line with the hierarchy pattern in the quark and charged lepton sector,
couplings to the first generation and mixing terms are generated by operators
of even higher order.  Operators of dimension 7 and 8 do not contribute to the
Yukawa matrix in our setup.  All sub-leading operators are at least of
dimension~9. The corresponding superpotential is:
\begin{align}
W_\text{sub} = & \frac{1}{M^6} \Bigl[
\left(\phi_{23}\,\overline{\phi}_{23}\right)^2
  (h\,\overline{\phi}_{23})(\phi_{23}\,\Psi_L\,\Psi_R) +
\left(\phi_{23}\,\overline{\phi}_3\right)^2
  (h\,\overline{\phi}_3)(\phi_{23}\,\Psi_L\,\Psi_R) \nonumber \\&+
\left(\phi_{23}\,\overline{\phi}_{23}\right)^3 (h\,\Psi_L\,\Psi_R) +
\left(\phi_{23}\,\overline{\phi}_3\right)^3 (h\,\Psi_L\,\Psi_R) \nonumber \\ &+
\left(\phi_{23}\,\overline{\phi}_3\right)^2 (\Psi_L\,\overline{\phi}_3)
(h\,\phi_{23}\,\Psi_R) +
\left(\phi_{23}\,\overline{\phi}_{23}\right)^2 (\Psi_L\,\overline{\phi }_{23})
(h\,\phi_{23}\,\Psi_R) \nonumber \\&+
\left(\phi_{23}\,\overline{\phi}_{23}\right)^2 (\Psi_R\,\overline{\phi}_{23})
(h\,\phi_{23}\,\Psi_L) +
\left(\phi_{23}\,\overline{\phi}_3\right)^2 (\Psi_R\,\overline{\phi}_3)
(h\,\phi_{23}\,\Psi_L)
\Bigr] \,.
\end{align}

In summary, we obtain the following parametrical structure of the low-energy
effective Yukawa matrices at leading order in $\epsilon$, neglecting all factors
of order unity:
\begin{equation}
 Y_{u/\nu} \approx \left( \begin{array}{ccc}
 0 &  \epsilon_{u}^3 &  \epsilon_{u}^3 \\
  \epsilon_{u}^3 &   \epsilon_{u}^2 &  \epsilon_{u}^2 \\
\epsilon_{u}^3 &  \epsilon_{u}^2 &   1
\end{array} \right) \epsilon_{u} \;,\;
 Y_{d/l} \approx \left( \begin{array}{ccc}
 0 &  \epsilon_{d}^3 &  \epsilon_{d}^3 \\
  \epsilon_{d}^3 &   \epsilon_{d}^2 &  \epsilon_{d}^2 \\
\epsilon_{d}^3 &  \epsilon_{d}^2 &   1
\end{array} \right) \epsilon_{d} \,.
\end{equation}

\subsection{Neutrino Masses and Lepton Mixing}
\label{sec:leptonmixing}

We now turn to the neutrino sector. The interplay between the Yukawa and
Majorana mass matrices  produces a non-hierarchical pattern in the PMNS neutrino
mixing matrix, along with the CKM quark mixing matrix as discussed above.

Assuming the quantum numbers given in Tab.~\ref{tab:fieldcontenttriplets}, we
write down operators that generate a right-handed neutrino Majorana mass, up
to dimension~12.
\begin{align}
W_\text{Maj}=&\, \frac{1}{M^9}\Phi _R^2 \Bigl[\
\left(\phi_3\,\overline{\phi}_{23}\right)^3
\left(\Psi_R\,\overline{\phi}_3\right)^2 +
(\phi_3\,\overline{\phi}_{23}) \left(\phi_3\,\overline{\phi}_3\right)^2
\left(\Psi_R\,\overline{\phi}_{23}\right)^2 +
\left(\phi_3\,\overline{\phi}_3\right)^3
\left(\Psi_R\,\overline{\phi}_3\right)^2
\nonumber \\ &+
(\phi_3\,\overline{\phi}_3) \left(\phi_3\,\overline{\phi}_{23}\right)^2
(\Psi_R\,\overline{\phi}_3) (\Psi_R\,\overline{\phi}_{23}) +
\left(\phi_3\,\overline{\phi}_{23}\right)^2
\left(\phi_{23}\,\phi_3\Psi_R\right)^2
\ \Bigr]\,.
\end{align}
Inserting the vacuum alignment and vevs given in (\ref{eq:vevstructure}), the
effective Majorana mass matrix takes the form
\begin{equation}
 M_{RR}=\left( \begin{array}{ccc}
 \sqrt[3]{\epsilon_\nu^{8}} & 0 & 0 \\
 0 & \epsilon_\nu^2  & \epsilon_\nu^2 \\
 0 & \epsilon_\nu^2  & 1 \\
\end{array} \right) \sqrt[3]{\epsilon_\nu^{8}}\; \frac{\vev{\Phi_R}^2}{M}\,.
\end{equation}

This Majorana mass matrix is diagonal up to corrections of order
$\epsilon_\nu^2$.  Furthermore, the right-handed neutrino mass eigenstates are
hierarchical $(\epsilon_\nu^3\,:\,\epsilon_\nu^2\,:\,1$).  If we allow for a
new expansion parameter $\epsilon_\nu\lesssim\epsilon_u^2$, we fulfill the
requirements for ``sequential dominance'', and the effective PMNS matrix
exhibits large mixing angles. For special choices of the order-one
coefficients, specific patterns (tri-bimaximal, golden-ratio) that fit actual
data are possible.

Concluding,  a model with flavour-triplet Higgses and flavour-triplet
flavons, supplemented by the extra symmetries listed in
Tab.~\ref{tab:fieldcontenttriplets}, is able to qualitatively explain
the observed SM flavour structure.

\section{Flavon Decuplet Model}
\label{sec:Decupletmodel}

As we have shown in the previous chapter, a model with only
flavour-triplet flavons can qualitatively reproduce the correct
structure of the fermion mass and mixing matrices.  However, to achieve this
success, we have to impose additional symmetries which eliminate terms that are
formally leading, but would generate an unwanted flavour structure. The
assignment of extra quantum numbers may be considered ad-hoc and unnatural.
Therefore, we should consider also non-fundamental flavon representations.

A generic Yukawa coupling takes the form $h\,\Psi_L\Psi_R$, in the
effective theory.  The prefactor of such a term, considered as a
symmetry-breaking spurion, has to transform according to one of the
terms in the reduction of the tensor product:
$\re{3}\otimes\re{3}\otimes\re{3}=\re{1}\oplus\re{8}\oplus\re{8}\oplus\re{10}$.

The simplest possibility for such a spurion is the vev of a single
flavon field, which has to transform according to either one
of the singlet, octet or anti-decuplet representations of $SU(3)_F$.
Further low-lying representations can occur only as a product. Therefore, the
sextet representation is not very promising, since it leads to similar
problems as the triplet case considered above.

The next larger representation of $SU(3)_F$ is the octet.  At each order in
the $1/M$ expansion, there are possible terms that generate a Yukawa coupling
once the octet vev is inserted.  However, if we explicitly construct the
invariant operators, we find no operator that singles out the top Yukawa
coupling. The resulting quark-mass pattern is very different from the observed
one, which should have a dominant entry for the third generation.  Hence, while
octet-induced structures may be present in the full theory, they should be
sub-dominant and merely act as perturbations to a Yukawa term with a different
origin.

The decuplet representation of $SU(3)_F$ is more promising.  It is a natural
choice, since it is the symmetric product of three triplets: $\re{10}_{ijk} =
\left(\re{3}_i \otimes \re{3}_j \otimes \re{3}_k \right)_\text{sym}$.  In the
following, we will investigate how far we can go by considering \emph{only}
decuplet flavons.  Just like in the triplet case, we will keep the discussion
on a semi-quantitative level and just determine hierarchy patterns that can
originate from such a model.  A full quantitative analysis would require a
detailed scan over a rather large parameter space, which we defer to future
work.

\subsection{Field content}

We study a model with the same basic symmetry (\ref{eq:gaugedsymmetry}) as in
the triplet-flavon case.  Matter and Higgs fields are in triplet
representations.  However, we put the flavon fields that are primarily
responsible for flavour-symmetry breaking, in $\re{10}$ and $\rb{10}$
representations of $SU(3)_F$.  The interactions that involve these fields
generate the desired flavour structure in a straightforward and simple way.

The breaking of $SU(3)_F$ proceeds in two steps by vevs of the
decuplet flavons $\phi_{i=2,3}$.  Supersymmetry requires D-term
flatness, so we also introduce their conjugate fields
$\overline{\phi}_i$. The latter will aid in generating the light
masses as subdominant contributions.  The PS-breaking multiplet
$\Phi_R$, which will be associated with neutrino-mass generation, is
also in the flavour triplet representation. The field content that is relevant 
for the flavour structure is shown in Tab.~\ref{tab:fieldcontentdecupletsneu}. 
Similar to the triplet model described above, we do not discuss the precise 
dynamics of PS and flavour-symmetry breaking, and omit extra fields that 
restore left-right symmetry and cancel gauge anomaly contributions.

\begin{table}[htbp]
\centering
  \begin{tabular}{ccccc}
  \hline \\[-2ex]
  Field		& $SU(3)_F$	& PS		& $Z_4$	& $Z_2^R$	\\[1ex]
  \hline \\[-1.5ex]
  $\Psi_L$	& $\re{3}$	& $(4,2,1)$		& $1$	& $-$ \\
  $\Psi_R$	& $\re{3}$	& $(\overline{4},1,2)$	& $1$	& $-$ \\
  $h$		& $\re{3}$	& $(1,2,2)$		& $1$	& $+$ \\[1ex]
  $\phi_3$	& $\rb{10}$	& $(1,1,1)$		& $1$	& $+$ \\
  $\phi_2$	& $\rb{10}$	& $(1,1,1)$		& $1$	& $+$ \\
  $\overline{\phi}_3$	& $\re{10}$	& $(1,1,1)$	& $3$	& $+$ \\
  $\overline{\phi}_2$	& $\re{10}$	& $(1,1,1)$	& $0$	& $+$ \\[1ex]
  $\Phi_R$	& $\rb{3}$	& $(4,1,2)$ 		& $1$	& $+$ \\[1ex]
  \hline
 \end{tabular}
 \caption{Transformation of the Higgs and matter and flavon superfields under
	  the gauge and flavour symmetries in the flavon decuplet model.}
 \label{tab:fieldcontentdecupletsneu}
\end{table}

The vev structure is chosen such that $\phi_3$ is responsible for the
third generation masses. $\phi_2$ gives masses to the second generation. The
mass of the light quarks and the Cabibbo angle are generated by higher
order corrections.  The vevs for the decuplets and anti-decuplets must
coincide, so SUSY is not broken via the D-terms. For simplicity we choose the
Higgs vev to be diagonal in flavour space. In summary, the vev structure is
given by:
\begin{subequations}
\begin{align}
 \vev{\phi_3}_{333} &= \vev{\overline{\phi}_3}_{333} \sim \epsilon \,,\\
 \vev{\phi_2}_{ijk} &= \vev{\overline{\phi}_2}_{ijk} \sim\epsilon^3 \quad
\text{with}\  i,j,k \geq 2 \,,\\[1ex]
\vev{h} &= (1,1,1)\,v_\text{MSSM} \,,\\
\vev{\Phi_R} &= (0,0,1)\,v_\Phi\,.
\end{align}
\end{subequations}

\subsection{Yukawa Potential}

Looking at the Yukawa potential, any model contains the term $Y_0$
(\ref{eq:trivinvariantstructure}) at dimension 3 which does not
involve flavons at all.   It can be removed by imposing a discrete
symmetry.  In the decuplet model, we choose a $Z_4$ symmetry.  This
will be the only extra symmetry that we need for a viable phenomenology.

The leading terms that do contribute have dimension four.  They
take the simple form (suppressing all flavour indices):
\begin{align}
  W_\text{lead} = \frac{y_1}{M}\, \phi_3\,\Psi_L\, \Psi_R\,h
                  + \frac{y_2}{M}\, \phi_2\,\Psi_L\, \Psi_R\, h\,.
\end{align}
The sub-leading terms of dimension five are given by
\begin{align}
 W_\text{dim5} = \frac{1}{M^2}\left[
  h\,\phi_3\,\overline{\phi}_2\,\Psi_L\,\Psi_R +
  h\,\phi_2\,\overline{\phi}_2\,\Psi_L\,\Psi_R
 \right]\,.
\end{align}
The superpotential which consists of these terms generates, after
flavour symmetry breaking, a Yukawa matrix consistent with the one
given in (\ref{eq:wantedYukmatrix}). As mentioned above, $\epsilon$ should
differ in the up and down sector.  We obtain
\begin{equation}
Y_{u/\nu} \sim
 \begin{pmatrix}
   0 & \epsilon_u^3 & \epsilon_u^3 \\
   \epsilon_u^3 & \epsilon_u^2 & \epsilon_u^2 \\
   \epsilon_u^3 & \epsilon_u^2 & 1
 \end{pmatrix}\epsilon_u\;,\qquad
Y_{d/l} \sim
 \begin{pmatrix}
   0 & \epsilon_d^3 & \epsilon_d^3 \\
   \epsilon_d^3 & \epsilon_d^2 & \epsilon_d^2 \\
   \epsilon_d^3 & \epsilon_d^2 & 1
 \end{pmatrix}\epsilon_d\,.
\label{eq:Yukawastructuredecuplet}
\end{equation}

For completeness, we list also the next order in the expansion, namely
the dimension-six operators
\begin{align}
 W_\text{dim6} = &\frac{1}{M^3}h\,\Psi_L\,\Psi_R \bigl[ \,
   \overline{\phi}_3\,\overline{\phi}_3\,\overline{\phi}_3 +
   \phi_3\,\phi_3\,\overline{\phi}_3   +
   \phi_2\,\phi_3\,\overline{\phi}_3 +
   \phi_2\,\phi_2\overline{\phi}_3 +
   \phi_3\,\overline{\phi}_2\,\overline{\phi}_2 +
   \phi_2\,\overline{\phi}_2\,\overline{\phi}_2 \, \bigr]\,.
\end{align}
This yields a subdominant contribution,
\begin{align}
 Y^f_\text{sub} \sim
 \begin{pmatrix}
   \epsilon^6 & \epsilon^8 & \epsilon^6 \\
   \epsilon^8 & \epsilon^6 & \epsilon^6 \\
   \epsilon^6 & \epsilon^6 & \epsilon^2 \\
 \end{pmatrix}\epsilon\,.
\end{align}
With suitable order-one coefficients inserted, we obtain a phenomenologically
viable flavour structure in the quark and charged lepton sector.

\subsection{Neutrino masses and Lepton mixing}

A neutrino mass and mixing structure that is consistent with neutrino
data can be generated by a see-saw mechanism with sequential dominance (cf.\
\cite{King:2003rf} and Sec.~\ref{sec:framework}).  The main
ingredient is a near-diagonal and hierarchical Majorana matrix for the
right-handed neutrinos.  In the current model, the relevant part of
the superpotential contains terms of dimension four and higher,
\begin{align}
 W_\text{Maj}\sim  &\frac{1}{M}\left(\Psi_R\,\Phi_R\right)^2
   \Bigl[ 1 +
   \frac{1}{M^2} \left( \phi_3\,\overline{\phi}_3 + \phi_2\,\overline{\phi}_3 +
     \overline{\phi}_2\,\overline{\phi}_2 \right) +
   \frac{1}{M^3} \left( \phi_3\,\overline{\phi}_2\,\overline{\phi}_3 +
     \phi_2\,\overline{\phi}_3 \overline{\phi}_2\right)
  \Bigr]\,.
\end{align}
We obtain a Majorana mass matrix for the right-handed neutrinos which
takes the form
\begin{align}
M_{\text{Maj}} \sim
\begin{pmatrix}
 \epsilon_\nu^6 & \epsilon_\nu^7 & \epsilon_\nu^5 \\
 \epsilon_\nu^7 & \epsilon_\nu^4 & \epsilon_\nu^4 \\
 \epsilon_\nu^5 & \epsilon_\nu^4 & 1 \\
\end{pmatrix} \;M_{N_R}\,.
\end{align}
This matrix is nearly diagonal and exhibits a hierarchical pattern with
eigenvalues of order $\epsilon_\nu^6\,:\,\epsilon_\nu^4\,:\,1$.  We need a
different expansion parameter $\epsilon_\nu<\epsilon_u$ in the neutrino sector
to fulfill the conditions for sequential dominance.  By inserting order-one
coefficients and allowing for CP-violating phases, we should be able to fit
flavour data, even if no additional terms and fields contribute.

\section{Conclusions}
\label{sec:conclusions}

We have studied supersymmetric models with a Pati-Salam gauge group and a
$SU(3)_F$ flavour symmetry where the Higgs bi-doublet occurs in three
generations and transforms as a flavour triplet.  The flavour symmetry is
spontaneously broken, presumably at high energy scales, and effective Yukawa
couplings are generated by a sequence of non-renormalizable interactions.
Associated with spontaneous flavour breaking, there are flavon fields.  The
low-energy phenomenology of quarks and neutrinos apparently favors flavons in
the decuplet representation of $SU(3)_F$, since few terms and a simple $Z_4$
extra symmetry qualitatively generate the observed hierarchies in quark and
neutrino data.

Since a qualitative description of flavour data is possible, a next step,
which we did not attempt in the present work, amounts to working out concrete
models and fit to the complete set of low-energy flavour data.  A specific
signature of the scenario is the presence of extra Higgs generations with
flavour quantum numbers.  A further study should address the Higgs potential
and any impact of Higgs-generation mixing on flavour data.  However,
gauge-coupling unification with intermediate left-right and Pati-Salam
symmetries is consistent with the presence of multiple Higgs generations at
low energies, so dedicated collider searches are definitely worthwhile.

\subsection*{Acknowledgments}

This work has been supported by a Coordinated Project Grant of the
Faculty of Science and Technology, University of Siegen.  We
acknowledge discussions with T.~Feldmann, T.~Mannel, K.~Schnitter,
C.~Luhn.

\end{document}